# Sub-cycle quantum electrodynamics in strongly laser-driven semiconductors

N. Tsatrafyllis[1], S. Kühn[2], M. Dumergue[2], P. Foldi[2,3], S. Kahaly[2], E. Cormier[2,4], I. A. Gonoskov[5], B. Kiss[2], K. Varju[2,6], S. Varro[2,7] and P. Tzallas[1,2*]

[1] *Foundation for Research and Technology-Hellas, Institute of Electronic Structure & Laser, PO Box 1527, GR-71110 Heraklion, Greece.*
[2] *ELI-ALPS, ELI-Hu Kft., Dugonics tér 13, H-6720 Szeged, Hungary*
[3] *Department of Theoretical Physics, University of Szeged, Szeged Hungary*
[4] *Univ Bordeaux, CNRS, CELIA, CEA, F-33405 Talence, France*
[5] *Max Planck Institute of Microstructure Physics, Weinberg 2, D-06120 Halle, Germany*
[6] *Department of Optics and Quantum Electronics, University of Szeged, Szeged Hungary*
[7] *Wigner Research Center for Physics, 1121 Budapest, Hungary*

[*]*Corresponding author e-mail: ptzallas@iesl.forth.gr*

Electrodynamical processes induced in complex systems like semiconductors by strong electromagnetic fields, have traditionally/conventionally been described using semi-classical approaches. Although these approaches, allowed the investigation of ultrafast dynamics in solids culminating in multi-petahertz electronics, they do not provide any access in the quantum optical nature of the interaction as they treat the driving-field classically and unaffected by the interaction. Here, using a full quantum-optical approach, we demonstrate that the sub-cycle electronic response in a strongly driven semiconductor crystal is imprinted in the quantum-state of the driving-field resulting in non-classical light-states carrying the information of the interaction. This vital step towards strong-field ultrafast quantum electrodynamics unravels information inaccessible by conventional approaches and leads to the development of a new class non-classical light sources.

The interaction of semiconductor crystals with strong electromagnetic fields has attracted keen interest over the last decade since it paves the way to combine the advantages of electronics and attosecond physics [1]. The fundamental mechanism underlying such interaction relies on the sub-laser-cycle electron dynamics induced in the crystal lattice by the strong driving laser-field. Untangling this sub-cycle dynamics led to pioneering discoveries in ultrafast solid state physics and attosecond optoelectronics [2, 3]. However, in all these studies the interaction is described using semi-classical theories treating the driving laser-field classically and unaffected by the interaction. Here, using a full quantum-optical approach, we reveal the sub-cycle quantum electrodynamics of the interaction and we show that the quantum state of the light exiting the crystal depicts non-classical features carrying the sub-cycle information of the interaction. This has been achieved using the high-harmonics (HH) generated by the interaction of crystalline Zinc-Oxide (ZnO) with intense mid-infrared (mid-IR) laser pulses.

According to semi-classical theories [1, 4-9], the generation of HH of order $q$ in the strong-field region is described via interband transitions using the concept of the electron recollision process in strong-field laser-atom interactions (Fig. 1a, b). Briefly, when an intense linearly polarized laser-field (with photon energy much lower than the energy band-gap of the material) interacts with a crystal, the electron escapes the valence band to the conduction band, it accelerates in the conduction band (likewise the hole in the valence band) gaining energy from the driving-field and recombines with the hole (in a multicenter recombination process along the crystal lattice) within the same cycle of the driving-field. This recombination process generates radiation with frequencies higher than the driving-field and, for multi-cycle interactions, leads to the emission of HH in the visible/UV region.

To unravel the influence of this process on the state of the driving laser-field, we have calculated and measured the photon number distribution of the mid-IR laser-field exiting the crystal by utilizing the UV/mid-IR anti-correlation approach (named hereafter "quantum spectrometer") that has been demonstrated for gases [10]. To gain insight on observation we have developed a model (see part 1 of Supplementary Materials) which describes the interaction using a quantized mid-IR laser-field. Figure 1a shows a schematic of the interaction. Before the interaction, the laser-field is in a coherent state [11, 12]. In the high-photon number limit, the probability distribution of this state (resulted by its projection to the photon number state) has a Gaussian form with its width reflecting the quantum noise of the light. A way to modify the quantum state of the light and its photon number distribution is to use a process which depends non-linearly on the instantaneous field amplitude [13-19]. In our case, the sub-cycle non-linear process which leads to HH generation (Fig. 1b), changes the quantum state of the driving-field and results to the generation of non-classical light-states having the photon number distribution shown in Fig. 1c. The distribution consists of a series of peaks which correspond to the HH spectrum (Fig. 1d). The peak structure appearing in the distribution, is a quantum-optical effect which cannot be explained by semi-classical theories. It results from quantum interference of the mid-IR photons absorbed (with the periodicity of the laser-field) towards the UV emission and the projection of the final light-state (generated with each shot) on a photon number state [13].

To measure the photon number distribution, we have used the experimental set-up shown in Fig. 2a (see part 2 of Supplementary Materials). A laser system which delivers few-cycle mid-IR pulses at 100 kHz repetition rate was used [20]. A linearly polarized laser pulse ($IR_0$) of ≈ 3.1 μm carrier wavelength and ≈ 75 fs duration, was focused into a 500 μm thick ZnO crystal where the harmonics were generated. For the measurement of the mid-IR photon statistics, we have used an $IR_0$ beam with photon number $N_{IR_0} \approx 4 \times 10^{14}$ photons per pulse and intensity on the crystal $I_{IR_0} \approx 5 \times 10^{11}$ W cm$^{-2}$. The generated harmonic spectrum (Fig. 2b) was recorded by a conventional spectrometer, while the photomultiplier (PMT) integrates the photon number of the harmonics generated mainly by interband transitions [6, 7] (i.e. with photon energy > 3.2 eV). The mid-IR beam exiting the crystal ($IR_1$), after being attenuated by a factor of $A \approx 3 \times 10^5$, was recorded by the photodiode $PD_s$. The signals of $PD_r$ ($S_{PDr}$), $PD_s$ ($S_{PDs}$) and PMT ($S_{PMT}$) were simultaneously recorded for each laser shot. $S_{PDs}$ was used for recording the photon number distribution of the mid-IR beam exiting the crystal, $S_{PDr}$ for reducing the energy fluctuations of the laser and $S_{PMT}$ to correlate the harmonic signal with that of the $S_{PDs}$ and remove the unwanted background caused by processes irrelevant to the harmonic generation.

A key step towards the realization of the quantum-optical approach is the verification of the energy conservation and the non-linearity of the interaction process. This was achieved by measuring the dependence of the missing mid-IR photon number $\Delta N_{IR} = N_{IR_0} - N_{IR_1}$ (where $N_{IR_1} = A \cdot N_{IR_s}^{(PD_s)}$ and $N_{IR_s}^{(PD_s)}$ is the mid-IR photon number exiting the crystal and reaching $PD_s$, respectively) and the generated UV photon number ($N_{HH}$ measured by the PMT) at the exit of the crystal on $I_{IR_0}$ (Fig. 3a). It is evident that both $\Delta N_{IR}$ and $N_{HH}$ have the same non-linear dependence on $I_{IR_0}$. After an increase with slope of $\approx 6 \pm 1$ they saturate at $I_{IR_0}^{(sat)} \gtrsim 5 \times 10^{11}$ W cm$^{-2}$. This is in agreement with the dependence of the signal of the individual harmonics (measured by the conventional spectrometer) on $I_{IR_0}$ in the non-perturbative regime (Fig. 3b). The above clearly reflects the energy conservation and depicts that both, $\Delta N_{IR}$ and $N_{HH}$, originate from the same non-linear interaction process.

The measured photon number distribution (red-line in Fig. 4a) is centered at $N_{IR_s}^{(PD_s)} \approx 8.70 \times 10^8$ photons per pulse and contains a series of confined peaks having a spacing of $\Delta N_{(IR_{PD_s})}^{(q)} = N_{(IR_{PD_s})}^{(q-1)} - N_{(IR_{PD_s})}^{(q)} \equiv \Delta N_{(IR)}^{(q)}/A \approx 6 \times 10^6$ photons per pulse. The distribution, named "mid-IR harmonic spectrum", reveals the HH spectrum as: (a) it exhibits a characteristic plateau and cut-off region; (b) the value of $\Delta q$ measured by the "mid-IR harmonic spectrum" (see parts 3 and 4 of Supplementary Materials) is in agreement with the value expected from the process leading to the generation of the odd and even harmonics i.e. $\Delta q \approx 1$; (c) is in agreement with the part of HH spectrum (with $q > 4$) obtained by the conventional spectrometer (blue-filled area in Fig. 4a) taking into account the absorption coefficient ($\alpha$) of ZnO crystal [21, 22] (black solid line in Fig. 4c). The above was confirmed by measurements performed when the ZnO crystal optical axis is perpendicular to the laser polarization (Fig. 4b). In this case only the odd harmonics are emitted (see part 5 of Supplementary Materials) [23] and the measured $\Delta q$ is found to be $\approx 2$. The latter constitutes the absolute validation check that the "mid-IR harmonic spectrum" reveals the HH spectrum.

The "mid-IR harmonic spectra", were used to measure the absorption coefficient of the crystal and determine the contribution of the different mechanisms participating in the HH generation process. Assuming that there are no data available about the absorption coefficient, its values can be obtained by dividing the harmonic signal of the conventional spectra (Fig. 2b and Fig. S5 of Supplementary Materials) with the corresponding $P_n$, the peak-normalized distribution of the "mid-IR harmonic spectra" i.e. $\alpha = S_{HH}/P_n$. The results (red points and blue rhombs in Fig. 4c) are found to be in fair agreement with the values obtained by conventional field free approaches (black solid line in Fig. 4c). However, the importance of this measurement relies on the capability of the approach to provide direct access to studies of optical properties of solids in the presence of strong fields [24].

To obtain the information about the mechanisms participating in the harmonic generation process, we compared the "mid-IR harmonic spectra" with the conventional harmonic spectra that has been corrected for the absorption $S_{HH}^{(\alpha)}$. Figure 4d shows the ratio of the peaks of the "mid-IR harmonic spectra" with the corresponding harmonic peaks of the conventional spectra ($P_n/S_{HH}^{(\alpha)}$). While both are in agreement for harmonics with $q > 4$ ($P_n/S_{HH}^{(\alpha)} \sim 1$) they depict a distinct difference for harmonics with $q < 5$ ($P_n/S_{HH}^{(\alpha)} \sim 10^{-4}$). In this part of the spectrum, $P_n$ drops by more than an order of magnitude compared to the plateau harmonics, while $S_{HH}^{(\alpha)}$ in the conventional spectra increases by more than an order of magnitude. This is because the "mid-IR harmonic spectrum" results from the anti-correlation with the harmonics generated in the non-perturbative regime, a procedure (not achievable by conventional spectrometers) that eliminates the detection of harmonics generated by other processes like multiphoton absorption (see part 2 of Supplementary Materials). The above is further supported by recent theoretical models [7] where the harmonic spectra which have been calculated using interband transitions are in fair agreement with the "mid-IR harmonic spectra" shown in Fig. 4a, b.

Concluding, we demonstrate that strongly laser-driven semiconductors lead to the generation of non-classical light-states having sub-cycle electric field fluctuations that carry the information of the sub-cycle dynamics of the interaction. These states have been used to recover the high-harmonic spectrum, resolve the mechanisms participating in the high-harmonic emission and measure the absorption coefficient of the crystal in the presence of strong laser-fields. The work connects the quantum optics and ultrafast electronics and paves the way for the development of a new class of compact squeezed light sources advancing studies ranging from quantum communication/information/computation to high precision interferometry applied for the detection of gravitational waves [19, 25].

**References**

1. A. Yu. Kruchinin, F. Krausz, V. S. Yakovlev, *Rev. Mod. Phys.* **90**, 021002 (2018).
2. T. T. Luu, M. Garg, S. Yu. Kruchinin, A. Moulet, M. Th. Hassan, and E. Goulielmakis, *Nature* **521**, 498 (2015).
3. M. Schultze, *et al. Nature* **493,** 75–78 (2012).
4. S. Ghimire, *et al.*, *Nat. Phys.* **7**, 138-141 (2010).

**Acknowledgments**

This work was supported by the: LASERLAB-EUROPE (no. 284464, EC's Seventh Framework Program), NFFA-Europe (Nr.: 654360), "HELLAS-CH" (MIS 5002735) funded by the Operational Program (NSRF 2014-2020) and co-financed by Greece and the European Union, European Union's Horizon 2020 research and innovation program MEDEA (Marie Sklodowska-Curie grant agreement no. 641789), European Social Fund (EFOP-3.6.2-16-2017-00005). ELI-ALPS is supported by the European Union and co-financed by the European Regional Development Fund (GINOP-2.3.6-15-2015-00001). E.C. acknowledge support from the Institut Universitaire de France. We thank K. Osvay for his support during the experimental campaign at ELI-ALPS and D. Charalambidis for fruitful discussions.

N. T. was the main contributor in the experiment runs and data analysis; S. Kü., M. D., S. Ka., E. C. contributed to the experiment, data analysis and manuscript preparation; P. F., S. V. developed the theoretical model; B. K. was responsible for the operation of the mid-IR laser system; K. V., I. A. G., contributed to the theoretical description of the HHG process; P.T. conceived and supervised the project, designed the experiment, and contributed to all aspects of the work.

**Figure captions**

FIG. 1. (a) A schematic of the interaction of the mid-IR pulses with the ZnO crystal. The width of the mid-IR field (pink-shaded area) before and after the interaction reflects the electric field variance ($\Delta E$) as has been calculated by the theoretical model. The blue-cylinder shows the HH beam after the crystal. (b) A simplified schematic representation of the multiphoton and interband transitions in a periodic potential induced by an intense mid-IR laser-field. (c) Distribution of the probability to absorb mid-IR photons towards harmonic emission ($P_n^{(abs)}$) calculated using the theoretical model. (d) Calculated HH spectrum.

FIG. 2. (a) Experimental set-up. M: Mid-IR plane mirrors. Polarizer: A system of two polarizers used to control the laser pulse energy. BS: Beam sampler used to reflect a small portion of the mid-IR beam towards the photodiode $PD_r$. $SM_1$: Spherical focusing mirror of 50 cm focal length. $IR_0$: Interacting Mid-IR beam. $IR_r$: Reference mid-IR beam. $IR_1$: Mid-IR beam exiting the crystal. $IR_s$: Mid-IR beam reaching the photodiode $PD_s$. $SM_2$: Spherical focusing mirror used to collect the mid-IR beam on $PD_s$. HH: High-harmonic beam. Si: Silicon plates used to reflect the harmonics towards the photomultiplier (PMT) and the conventional spectrometer (C-Spec). F: Neutral density filters used for the attenuation of the mid-IR beam. (b) High-harmonic spectrum at the exit of the crystal recorded with a conventional spectrometer at $I_{IR_0} \approx 5 \times 10^{11}$ W cm$^{-2}$ when the optical axis of the crystal is parallel to the laser polarization.

FIG. 3. (color online) (a) Dependence of the missing mid-IR photon number $\Delta N_{IR} = N_{IR_0} - N_{IR_1}$ (red points) and the generated UV photon number $N_{HH}$ obtained by the PMT (blue points) at the exit of the crystal on $I_{IR_0}$. The upper x-axis shows the photon number $N_{IR_0}$ of the $IR_0$ beam. (b) Dependence of the harmonic signal on $I_{IR_0}$ recorded by the conventional spectrometer. The black-dashed lines show the linear fit on the raw data. The slopes depict the order of non-linearity. The dashed red line shows the value of the $I_{IR_0}$ used for the measurements of the mid-IR photon number distribution.

FIG. 4. (color online) (a) and (b) "Mid-IR harmonic spectra" and the conventional HH spectra recorded when the optical axis of the crystal is parallel and perpendicular to the laser

polarization, respectively. In (a) and (b), the red lines show the photon number distribution of the mid-IR beam (IR$_s$) reaching the detector PD$_s$. The blue filled areas show the conventional HH spectra at the exit of the crystal obtained after taking into account the absorption coefficient of the crystal. (c) Measurement of the ZnO absorption coefficient. The black line shows the absorption coefficient of the ZnO crystal measured by conventional approaches. (d) Measurement of the contribution of the different harmonic generation mechanisms.

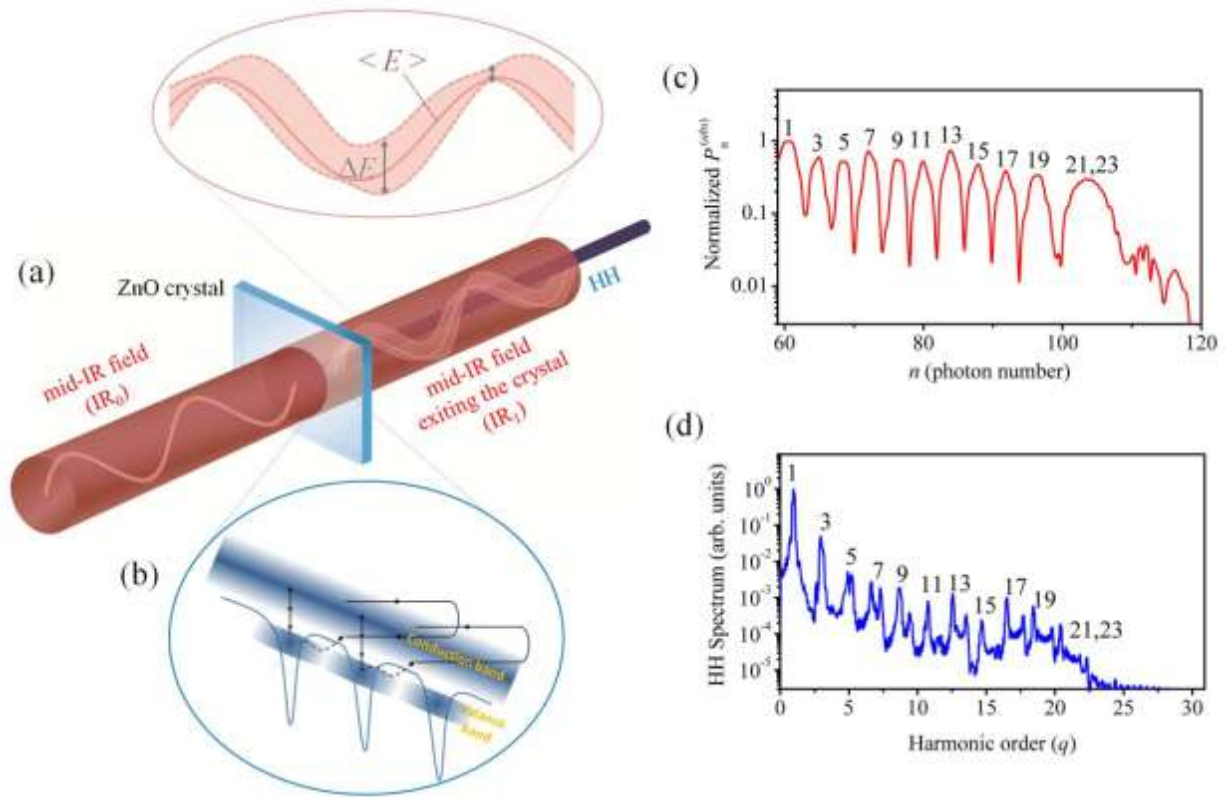

**Figure 1.**

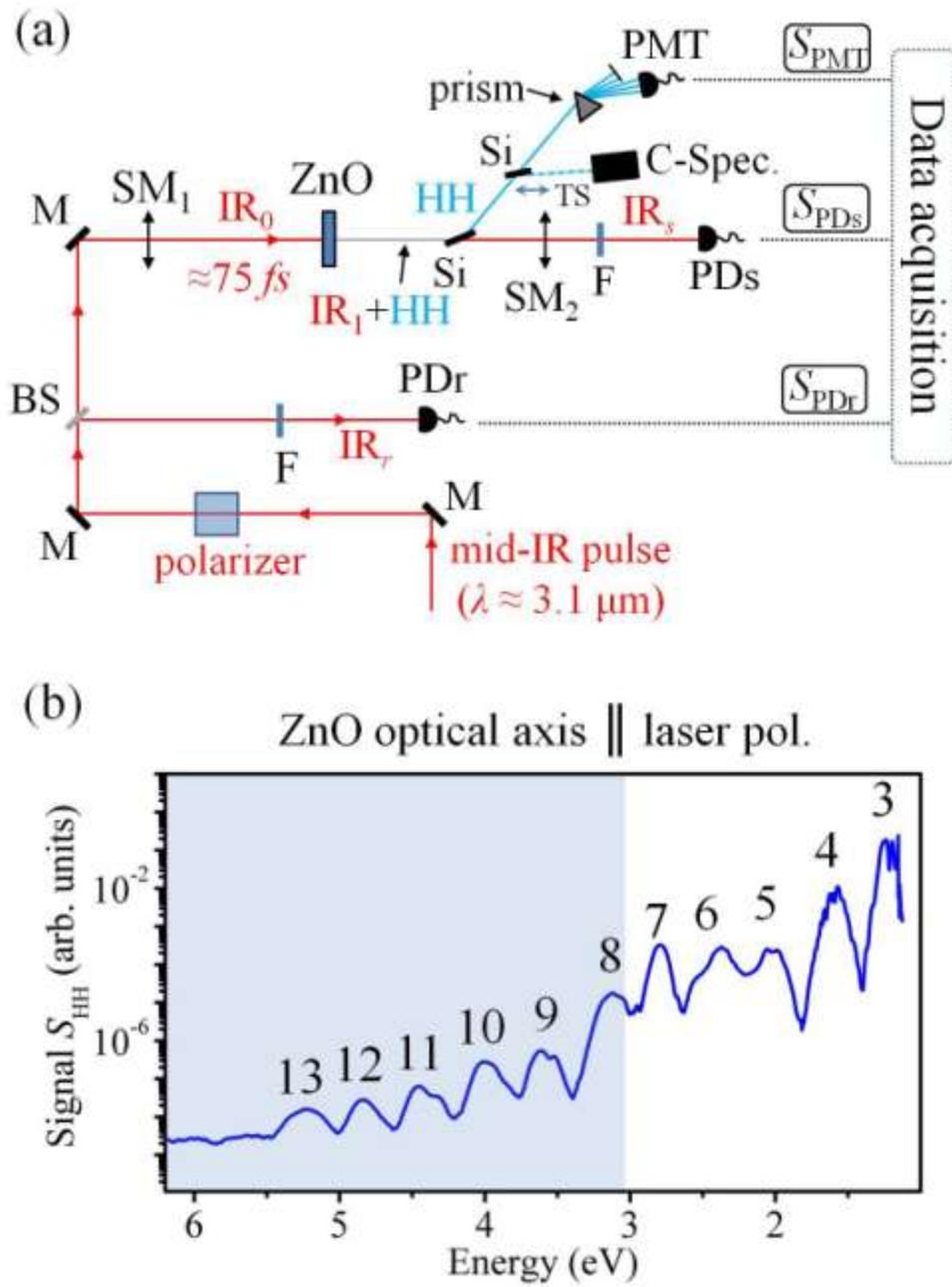

**Figure 2.**

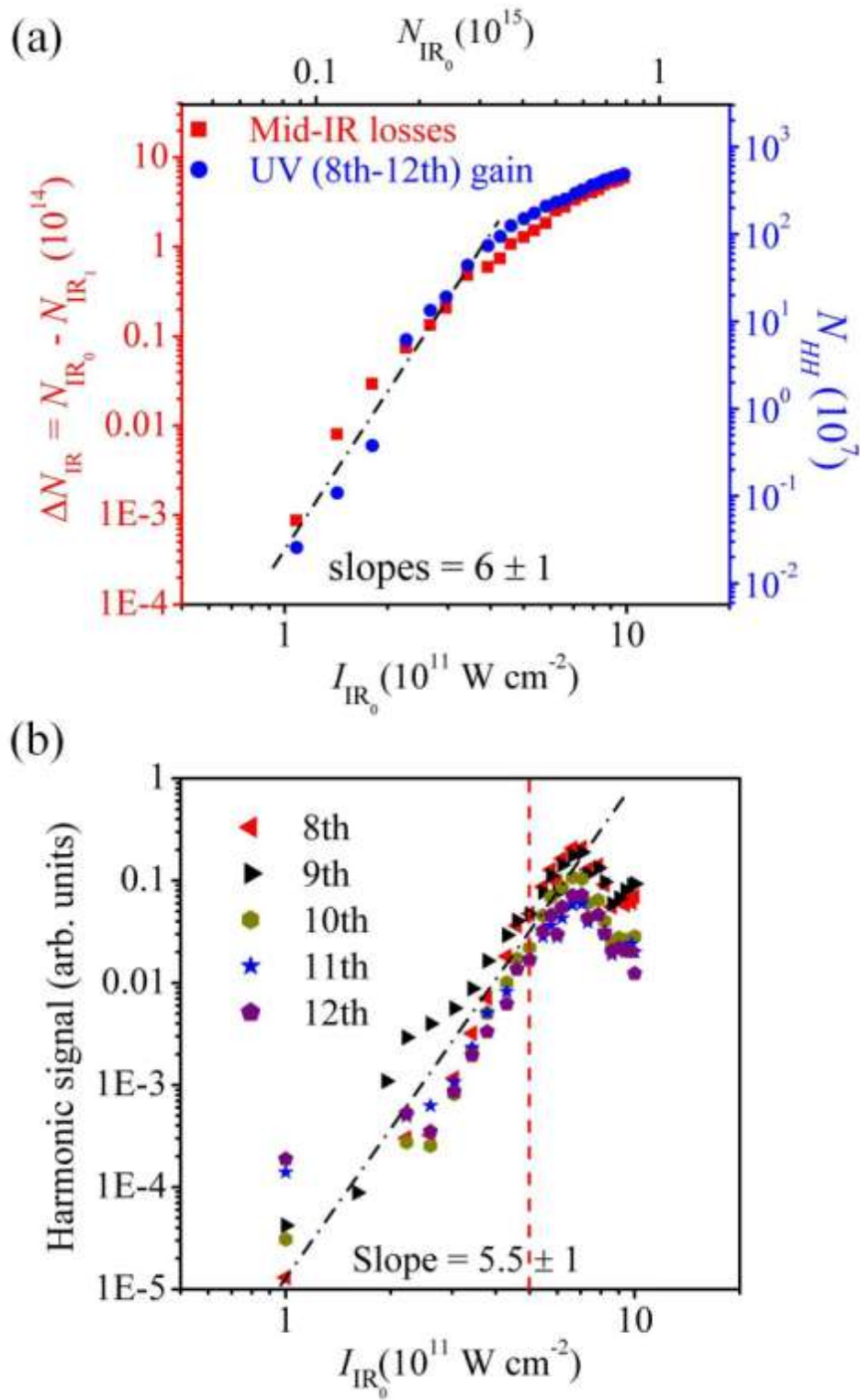

**Figure 3.**

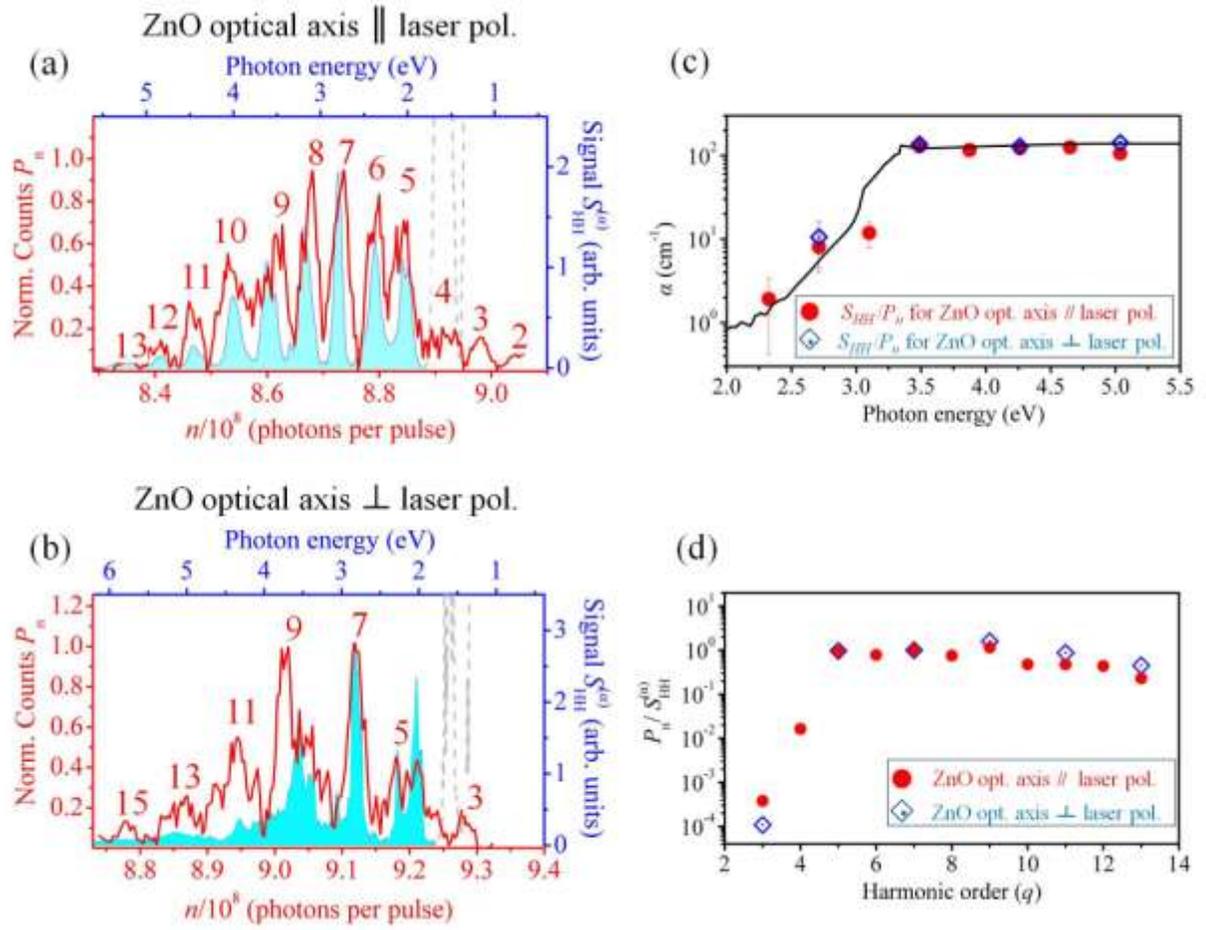

**Figure 4.**

# Supplementary Material

**Part 1: Theoretical model used to calculate the mid-IR photon number distribution at the exit of the crystal.**

The Hamiltonian which describes the interaction of a bulk crystal with a quantized single-mode of mid-IR electromagnetic radiation, in dipole approximation, is given by

$$H = \hbar\omega_L \left(\alpha_L^+ \alpha_L + \frac{1}{2}\right) + \frac{1}{2m_e}[\boldsymbol{p} - e\boldsymbol{A}]^2 + U(\boldsymbol{r}), \tag{1}$$

where $e$ and $m_e$ are the charge and the mass of the electron, respectively (where electron-electron interactions are neglected). The first term in eq. (1) describes the free electromagnetic radiation, the second one means light-matter coupling (in velocity gauge, $A = \sqrt{\hbar/2\omega_L \varepsilon_0 V}(\alpha_L + \alpha_L^+)$, with $\alpha_L^+$ and $\alpha_L$ being the creation and annihilation operators and $V$ is the quantization volume), while $U(\boldsymbol{r})$ is the lattice periodic potential of the crystal at the electron position $\boldsymbol{r}$.

In case of HH generation, the quantized HH modes (indexed by $q$) also appear in the Hamiltonian:

$$H = \hbar\omega_L \left(\alpha_L^+ \alpha_L + \frac{1}{2}\right) + \sum_q \hbar\omega_q \left(\alpha_q^+ \alpha_q + \frac{1}{2}\right) + \frac{1}{2m_e}[\boldsymbol{p} - e\boldsymbol{A}]^2 + U(\boldsymbol{r}) \tag{2}$$

In this case the vector potential $A$ contains contributions from all (mid-IR and HH) modes. For a given crystal structure (i.e. potential $U(\boldsymbol{r})$), the eigenstates of the field-free Hamiltonian $H_0 = \frac{p^2}{2m_e} + U(\boldsymbol{r})$ can be obtained in the form of Bloch-states $|k\rangle = e^{ikr} u_k^m(\boldsymbol{r})$, where the functions $u_k^m(\boldsymbol{r})$ are lattice-periodic, with $m$ denoting the band index. The crystal momentum dependent eigenenergies $E^m(\boldsymbol{k})$ provide the dispersion relation. The Hamiltonian of eq. (1) induces "vertical" transitions only, i.e. it does not mix states with different crystal momenta $\boldsymbol{k}$. The undisturbed crystal is assumed to be in thermal equilibrium, while initially the mid-IR field can be described by a coherent state $|\alpha\rangle$, and the HH modes are in their zero-photon states $|n_q = 0\rangle$. That is, the initial state of the total system is given by a quantum mechanical density matrix

$$\varrho(0) = \sum_{k,m} f(E^m(\boldsymbol{k}), \mu, T) |\psi\rangle\langle\psi|, \quad |\psi\rangle = |k\rangle|\alpha\rangle|\{0\}\rangle, \tag{3}$$

where the summation runs over the first Brillouin zone, $f$ denotes the Fermi distribution (with $\mu$ being the chemical potential and $T$ is the temperature) and the explicit tensorial notation has been omitted. The time evolution of $\varrho$ is governed by the von Neumann equation, $i\frac{d}{dt}\varrho(t) = [H, \varrho(t)]$, where a commutator appears on the right hand side. Having obtained $\varrho(t)$, the complete coupled dynamics of the system (crystal + mid-IR-mode + HH-modes) are known, and the expectation values of all relevant operators can be calculated. For the photon number operators of the exciting field and the $q$th HH mode we have: $\langle N_L \rangle = \langle a_L^+ a_L \rangle = \text{Tr}[\varrho(t)\, a_L^+ a_L]$ and $\langle N_q \rangle = \text{Tr}[\varrho(t)\, a_q^+ a_q]$, respectively.

However, numerical limitations force us to simplify the approach outlined above. In order to keep the essence of the model but reducing its dimensionality, we restrict our calculations to one spatial dimension (1D), and take only two HH modes into account. An 1D model potential $U(r)$ is considered, and the Bloch-states together with the dispersion relation $E^m(k)$ are obtained numerically. In order to comply with the experiment, a potential has been chosen, for which the band gap $\Delta = \min(E^c - E^v)$ is 3.2 eV. Furthermore, we took only two bands (the valence band $v$, and the first conduction band $c$) into account. For the actual value of $\Delta$ at room temperature $f(E^v(\mathbf{k}))$ and $f(E^c(\mathbf{k}))$ can be approximated by 1 and 0, respectively. With these initial conditions, the dynamics are calculated numerically, using a finite coherent-state basis [1] for the exciting field, a number-state basis for the HH modes and a Bloch-state basis for the crystal. The matrix elements of the operator $p$ together with their phase were also obtained numerically [2]. Additionally, since the dynamics are dominated by the action of the strong driving-field, we assumed that $c$-$v$ transitions corresponding to different values of $k$ are independent in the sense that the back-action of a given transition on the exciting mode negligibly modifies the dynamics of the others. Using these assumptions, an appropriate Cash-Karp Runge-Kutta routine was used to compute the time evolution so that the quantum mechanical norm is preserved within a relative error of $10^{-7}$.

In order to obtain the time resolved details of the HH process, we have calculated the back action of the electron motion in the crystal lattice on the driving-field (Fig. S1a,b) taking into account both inter- and intraband contributions [3]. For reasons of convenience the calculation was performed using a 7-cycle (which corresponds to the pulse duration used in the experiment) single mode mid-IR laser-field with initial photon number $N_{IR_0} = 70$ photons.

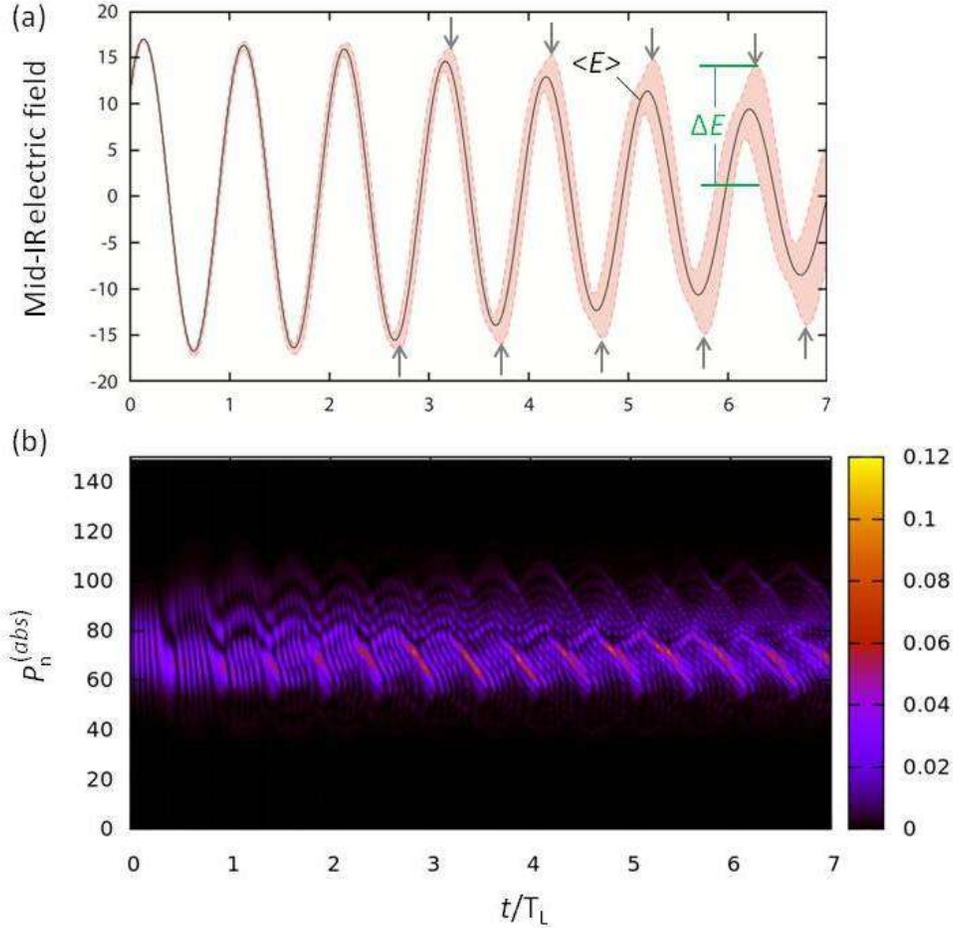

Fig. S1. (a) Calculated mid-IR electric field during the interaction with the ZnO crystal when the polarization of the driving-field is perpendicular to the crystal optical axis. The expectation value of the electric field ($<E>$) is shown with black solid line and its variance ($\Delta E$) with pink shaded area. The gray arrows show the time intervals where $\Delta E$ depicts a pronounce distortion. (b) Distribution of the probability to absorb mid-IR photons towards UV emission ($P_n^{(abs)}$) during the interaction with the crystal. $T_L$ is the period of the mid-IR field. Note that $\Delta E$ is not related directly to the width of the distribution $P_n^{(abs)}$.

The intensity of the field in the interaction region was artificially kept in the strong-field regime in order to be able to produce the high-order harmonics shown in Fig. 1d of the main text of the manuscript. These parameters simplify the numerical calculations and allow a qualitative description of the process. Note that, however, using photon number as high as in the experiment, the results point towards the same direction. Figure S1b shows the calculated expectation value of the electric field ($<E>$) (black solid line) and its variance ($\Delta E$) (pink shaded area) during the interaction with the medium. The reduction of the expectation value of the electric field is mostly related with the mid-IR photons absorbed by the system without excluding the effect of the collapse of the quantum optical properties [4]. The sub-cycle influence of the HH generation process on the coherent state of the driving-field is shown in the values of $\Delta E$ which depict a pronounce distortion (shown with gray arrows) every half cycle of the field. The broadening of $\Delta E$ (although is influenced by the reduction of $<E>$) is mainly associated with the harmonic generation process and will be discussed in detail in the following paragraph.

The influence of the electron motion in the driving-field is more pronounced in Fig. S1b which shows the distribution of the probability to absorb mid-IR photons towards UV emission ($P_n^{(abs)}$). The distribution depicts the well known features observed for the HH spectra recorded by conventional approaches using few- and multi-cycle laser-fields. In the 1st half cycle of the field the probability to absorb mid-IR photons towards UV emission is confined around the continuum photon number distribution of the coherent state. In the 2nd half cycle $P_n^{(abs)}$ depicts a continuum structure significantly broader than the distribution of the 1st half cycle. After the 2nd half-cycle the width of $P_n^{(abs)}$ remains approximately constant. After the 3rd half cycle and up to 3rd - 4th cycle of the field, $P_n^{(abs)}$ depicts a quasi-continuum "multi-peak" structure with the position of the "peaks" to be time dependent within the half cycles of the field. After the 4th-5th cycle, the distribution is approximately constant and depicts a multi-peak structure. Finally, the distribution of the field exiting the medium (resulting by the integral of $P_n^{(abs)}$ over the last cycle of the interaction, in order to incorporate any CEP instability of the field at the exit of the medium) depicts the multi-peak structure (with $\Delta q \approx 2$ as it is expected from the generation of odd harmonics) which shown with black line in Fig. S2a. The green line in Fig. S2a is the background distribution which contains the probability to absorb mid-IR photons towards processes other than the harmonic emission. The red line in Fig. S2b (which is shown in log scale in Fig. 1c of the main text of the manuscript) shows the distribution resulted after subtracting the background from the $P_n^{(abs)}$ (shown with black line in Fig. S2a). Concluding, it is found that for multi-cycle interactions, due to quantum interference effects taking place periodically with the laser-field, $P_n^{(abs)}$ of the field exiting the medium depicts a multi-peak structure corresponding to the HH.

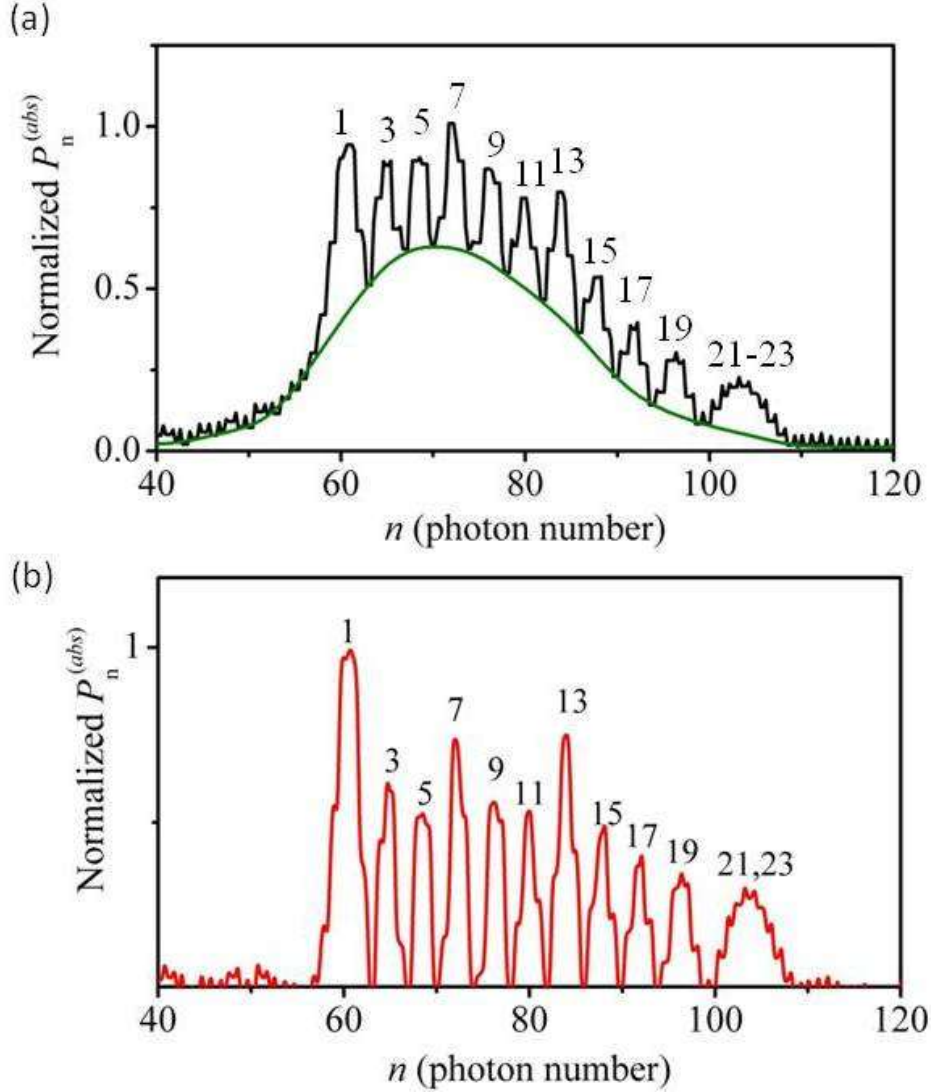

Fig. S2. (a) Black line: $P_n^{(abs)}$ of the field exiting the medium. Green line: background distribution which contains the probability to absorb mid-IR photons towards processes other than the harmonic emission. (b) Distribution resulted by subtracting the background from the $P_n^{(abs)}$ (shown with black line in Fig. S2a).

We note, that the number of mid-IR photons corresponding to the $q$th harmonic peak in the distribution of Fig. 1c of the main text of the manuscript, does not lead to the production of a single $q$th harmonic. The peak structure appearing in the distribution, is a quantum optical effect which cannot be explained by semi-classical theories. It results from quantum interference of the mid-IR photons absorbed (with the periodicity of the laser-field) towards the UV emission and the projection of the final light-state (generated with each shot) on a photon number state. In other words, in a multi-shot experiment, while each laser shot produces all the harmonics the histogram describing the photon statistics of the mid-IR field depicts a multi-peak structure.

To this end we would like to note that the present theoretical model provides only a qualitative information regarding the main features of the mid-IR photon number distribution and HH spectrum. The calculation of the detailed structure of the mid-IR photon distribution (like the broadening of the distribution, the relative amplitude of the "mid-IR harmonic" peaks, "mid-IR harmonic" peak structure, etc.) and HH spectrum (like harmonic peak structure, relative amplitude of the harmonic peaks, relative contribution of the multiphoton/intra- and inter-band transitions etc.) requires the consideration of the complete band structure of the crystal and the propagation effects of a multimode mid-IR field in the crystal, which is beyond the scope of this work.

**Part 2: Experimental approach**

The experiment was performed at ELI-ALPS using a laser system which delivers few-cycle mid-IR pulses at 100 kHz repetition rate with energy up to 150 μJ per pulse and shot-to-shot energy stability below 2 %. A linearly polarized laser pulse at ≈ 3.1 μm carrier wavelength, was passing through a system of two polarizers used to control the pulse energy and a beam sampler, which reflects a small portion of the beam towards the photodiode $PD_r$. The shot-to-shot energy stability of the laser after passing through the system of polarizers measured by the $PD_r$ was found to be close to 4 %. The transmitted beam ($IR_0$) was focused by a 50 cm focal length spherical mirror into a 500 μm thick ZnO crystal where the harmonics were generated. The duration of the pulse and the beam diameter on the crystal were ≈ 75 fs and ≈ 300 μm, respectively. In order to avoid saturation effects in the harmonic generation process and minimize the distortions of the driving-field during the propagation in the crystal, the photon statistics measurements were performed at a laser intensity of $I_{IR_0} \approx 5 \times 10^{11}$ W cm$^{-2}$. The generated harmonics are reflected by a Silicon (Si) plate towards a calibrated (with an uncertainty of a factor of ≈ 3) PMT which is used in the non-saturated region. A spectral filtering arrangement, consisting of a $CaF_2$ prism and an aperture, was placed before the PMT in order to select all the harmonics with photon energy > 3 eV. The generated harmonics were spectrally resolved by inserting in the harmonic beam path a second Si plate which reflects the harmonics towards a conventional spectrometer (C-spec). Both Si plates were placed at ≈ 78º angle of incidence in order to increase the reflectivity of the UV harmonics and reduce the contribution of visible/mid-IR radiation reaching the spectrometer. The mid-IR beam exiting the crystal, after passing through the first Si and neutral density filters was attenuated by a factor of $A \approx 3 \times 10^5$ and recorded by the photodiode $PD_s$ (e.g. when the ZnO crystal is out of the beam path the mid-IR photon number reaching the $PD_s$ is $N_{IR_s}^{(PD_s)} = N_{IR_0}/A \approx 1.3 \times 10^9$ photons per pulse). The signals of $PD_r$, $PD_s$ and PMT were simultaneously measured for each laser shot by a 12 bit data acquisition system. The data analysis process is extensively described in ref. [10] of the main text of the manuscript. Briefly, in the present work the photon number distribution was recorded keeping only the mid-IR laser pulses with energy stability at the level of 1 %. After that, we have created a UV/mid-IR joint distribution which contains $\approx 10^6$ uncorrelated values. The width of this distribution was found to be $\approx 5$ % of the mean value (Fig. S3).

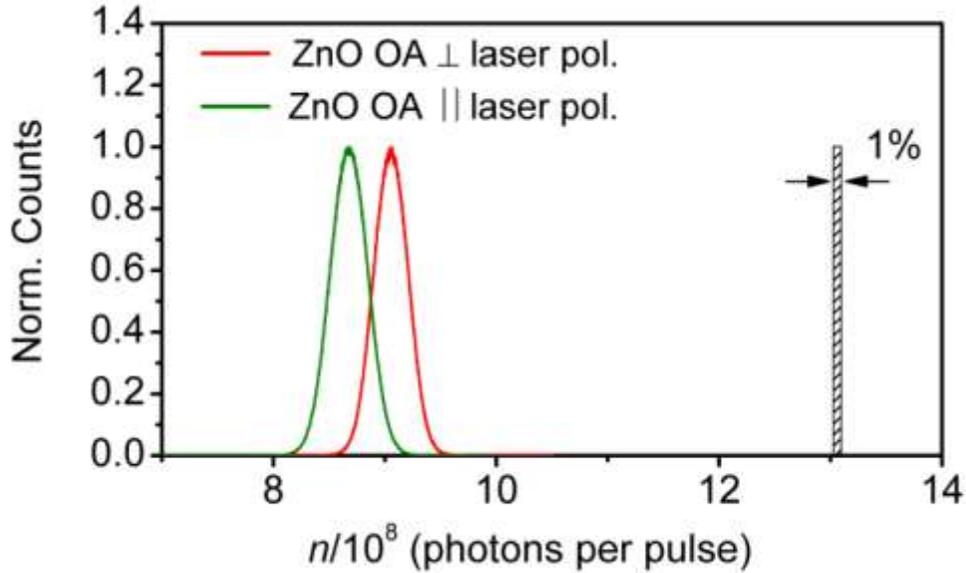

Fig. S3. Green line: Mid-IR distribution recorded by $PD_s$ when the ZnO crystal axis is parallel to the laser polarization after keeping only the mid-IR laser pulses with energy stability at the level of 1 % (black-line-shaded area). Red line: Mid-IR distribution recorded by PDs when the ZnO crystal axis is perpendicular to the laser polarization. The center of the distributions found to be at $8.70 \times 10^8$ photons per pulse and $9.07 \times 10^8$ photons per pulse for the ZnO optical axis parallel and perpendicular to the laser polarization, respectively.

This broadening (compared to the 1% laser distribution of the selected pulses) is associated with the non-linearity of the interaction which varies as $I_{IR_0}^6$. The center of the distribution (which is found to be at $8.70 \times 10^8$ photons per pulse and $9.07 \times 10^8$ photons per pulse for ZnO optical axis parallel and perpendicular to the laser polarization) is determined with an accuracy of $(5\%)/\sqrt{10^6} = 5 \times 10^{-3}$ %. This value defines the width of the anti-correlation diagonal which was used in order to obtain the mid-IR photon distributions shown in Fig. 4a, b and sets the upper limit in the resolution of the measurement. Additionally, the anti-correlation diagonal filters out the detection of processes with non-linearity different than 6. This is because the values of these processes in the UV/mid-IR joint distribution are lying outside the anti-correlation diagonal. For example, a joint distribution created by processes of different non-linearity (e.g. the missing mid-IR photons which varies as $I_{IR_0}^6$ and the 2nd (or 3rd) harmonic photons generated by multiphoton excitation and varying as $I_{IR_0}^2$ (or $I_{IR_0}^3$)) will be asymmetric resulting in a curve that deviates from the anti-correlation diagonal resulting from a symmetric joint UV/mid-IR distribution.

**Part 3: Calibration of the "mid-IR harmonic spectra"**

In the "mid-IR harmonic spectra" shown in Fig. 4a and 4b the values of $\Delta N_{(IR_{PD_S})}^{(q)}$ are found to be $\approx 6 \times 10^6$ and $\approx 9 \times 10^6$ photons per pulse, respectively. Thus, it is evident that, when the crystal optical axis is perpendicular to the laser polarization (i.e. only odd harmonics are generated) the spacing between the peaks is approximately double compared to the spacing of the peaks for crystal optical axis parallel to the laser polarization.

The calibration procedure of the "mid-IR harmonic spectra" described below, is commonly used for the calibration of the harmonic spectrum recorded by a conventional spectrometer i.e. a conventional harmonic spectrum can be calibrated by identifying the position of a single harmonic peak and then using the linear dependence of harmonic photon energy on $q$. The assignment of the peaks of the photon number distribution with harmonic order is based on the linear dependence of $N_{(IR_{PD_S})}^{(q)}$ on $q$ (taking into account the values of $\Delta N_{(IR_{PD_S})}^{(q)}$) and the first observable peaks in the distributions of Fig. 4a and 4b which correspond to the first harmonic with $q > 1$ i.e. 2$^{nd}$ and 3$^{rd}$ harmonic, respectively. This is because the peak which corresponds to $q = 1$ is not possible to be recorded with the present approach as these mid-IR photons are also detected by PD$_s$ and hence are lying outside the anti-correlation diagonal. The photon number value $N_{IR_S}^{(PD_S)}(q = 1)$ can be found by adding the measured $\Delta N_{(IR_{PD_S})}^{(q)}$ from the photon number value of the first observable peak in the distribution. In the "mid-IR harmonic spectra" shown in Fig. 4a and 4b these values are $\approx 9.04 \times 10^8$ photons per pulse and $\approx 9.30 \times 10^8$ photons per pulse. Consequently the $N_{IR_S}^{(PD_S)}(q = 1)$ in Fig. 4a and 4b is $\approx 9.1 \times 10^8$ photons per pulse and $\approx$ $9.4 \times 10^8$ photons per pulse, respectively. Also, by extrapolating to zero the linear dependence of $N_{(IR_{PD_s})}^{(q)}$ on $q$, it can be found that the value of $N_{IR_s}^{(PD_s)}(q=0)$ (which reflects the remained mid-IR photon number resulting from the absorption due to processes other than harmonic emission) coincide with the value of $q = 0$ obtained by the conventional harmonic spectra. The $N_{IR_s}^{(PD_s)}(q=0)$ in Fig. 4a and 4b was found to be $\approx 9.16 \times 10^8$ photons per pulse and $\approx 9.48 \times 10^8$ photons per pulse, respectively. The value of $\Delta q$ was extracted from the "mid-IR harmonic spectra" using the relation $\Delta q = q\Delta N_{(IR_{PD_s})}^{(q)}/(N_{IR_s}^{(PD_s)}(q=0) - N_{(IR_{PD_s})}^{(q)})$. In cases where $N_{IR_s}^{(PD_s)}(q=0)$ cannot be determined by using the first observable peaks (which might not be clearly detectable in a distribution) the assignment of the "mid-IR harmonic spectra" can be achieved by extrapolating to zero the linear dependence of $N_{(IR_{PD_s})}^{(q)}$ on $q$ using as a reference the photon number value which corresponds to characteristic harmonics in the spectrum like cut-off harmonics or harmonics with double peak structure (like the structure of the 5th harmonic shown in Fig. 4b).

**Part 4: Mid-IR photon number estimation**

For the estimation of the mid-IR photon number distribution we have used the measurements performed when the ZnO optical axis is parallel to the laser polarization. At the nominal intensity of $I_{IR_0} \approx 5 \times 10^{11}$ W cm$^{-2}$, $\Delta N_{IR} = N_{IR_0} - N_{IR_1}$ (where $N_{IR_1} = A \cdot N_{IR_s}^{(PD_s)}$ and $N_{IR_s}^{(PD_s)}$ is the mid-IR photon number exiting the crystal reaching the PD$_s$, respectively) and $N_{HH}$ were found to be $\approx 1.4 \times 10^{14}$ photons per pulse and $\approx 1.5 \times 10^9$ photons per pulse, respectively. We note that $\Delta N_{IR}$ is the sum of the mid-IR photons used for the harmonic emission ($N_{HH}^{(IR)} = \sum_q N_{(IR)}^{(q)}$ where $N_{(IR)}^{(q)}$ is the mid-IR photons absorbed for the $q$th harmonic emission) and all other processes ($N_{abs}^{(IR)}$) i.e. $\Delta N_{IR} = N_{abs}^{(IR)} + N_{HH}^{(IR)}$. $N_{HH}$ was obtained from the photon number measured by the PMT ($N_{HH}^{(PMT)}$) taking into account the $\approx 93\%$ energy losses (in the range around 3.8 eV) introduced mainly by the Si plate placed in the harmonic beam path i.e. $N_{HH} \approx 14 \times N_{HH}^{(PMT)}$. Taking into account the contribution of the harmonics 8th to 12th, the latter results that the photon number of the individual harmonics ($N^{(q)}$) is in the range from $10^8$ to $10^9$ photons per harmonic. Considering the absorption coefficient ($\alpha \approx 1.3 \times 10^2$ cm$^{-1}$ for $\hbar\omega > 3.2$ eV) of the ZnO crystal, it turns out that the measured harmonic photon number is reduced due to absorption effects in the medium by a factor of $B \approx 7\times10^2$. Considering, that $q$ mid-IR photons are required for the generation of the $q$th harmonic, it can be estimated that, for $q = 10$, the number of mid-IR photons absorbed towards harmonic emission $N_{(IR)}^{(q)} = B \cdot q \cdot N^{(q)}$, is in the range of $10^{12}$ to $10^{13}$ photons per harmonic and the photon number difference between the consecutive peaks ($\Delta q=1$) in the mid-IR probability distribution is $\Delta N_{(IR)}^{(q)} = B \cdot \Delta q \cdot N^{(q)}$ in the range of $10^{11}$ to $10^{12}$ photons per pulse. Consequently, the signal of PD$_s$ which is located around $N_{IR_s}^{(PD_s)} = N_{IR_1}/A \approx 8.70 \times 10^8$ photons per pulse is expected to have a value of $\Delta N_{(IR_{PD_s})}^{(q)} = N_{(IR_{PD_s})}^{(q-1)} - N_{(IR_{PD_s})}^{(q)} = \Delta N_{(IR)}^{(q)}/A$ being in the range of $\sim 10^6$ photons per pulse. This is found to be in fair agreement with the measurement. However, we note that the estimation of the values of $N^{(q)}$, $N_{(IR)}^{(q)}$, $\Delta N_{(IR)}^{(q)}$ and $\Delta N_{(IR_{PD_s})}^{(q)}$ given above is rough, and performed only to depict the feasibility of performing mid-IR photon number distribution measurements for revealing the high-harmonic spectrum.

**Part 5: Conventional high-harmonic spectrum at the exit of the spectrometer.**

Figure S4a and S4b show the harmonic spectra as were recorded by the conventional spectrometer when the optical axis of the crystal was parallel and perpendicular to the laser polarization, respectively.

In order to obtain the spectrum at the exit of the crystal, the spectra have been divided by the spectral response of the spectrometer (black line in Fig. S4c) and the reflectivity of the two Si plates (blue line in Fig. S4c). The results for laser polarization parallel and perpendicular to the crystal optical axis are shown in Fig. 2b of the main text of the manuscript and in Fig. S5, respectively.

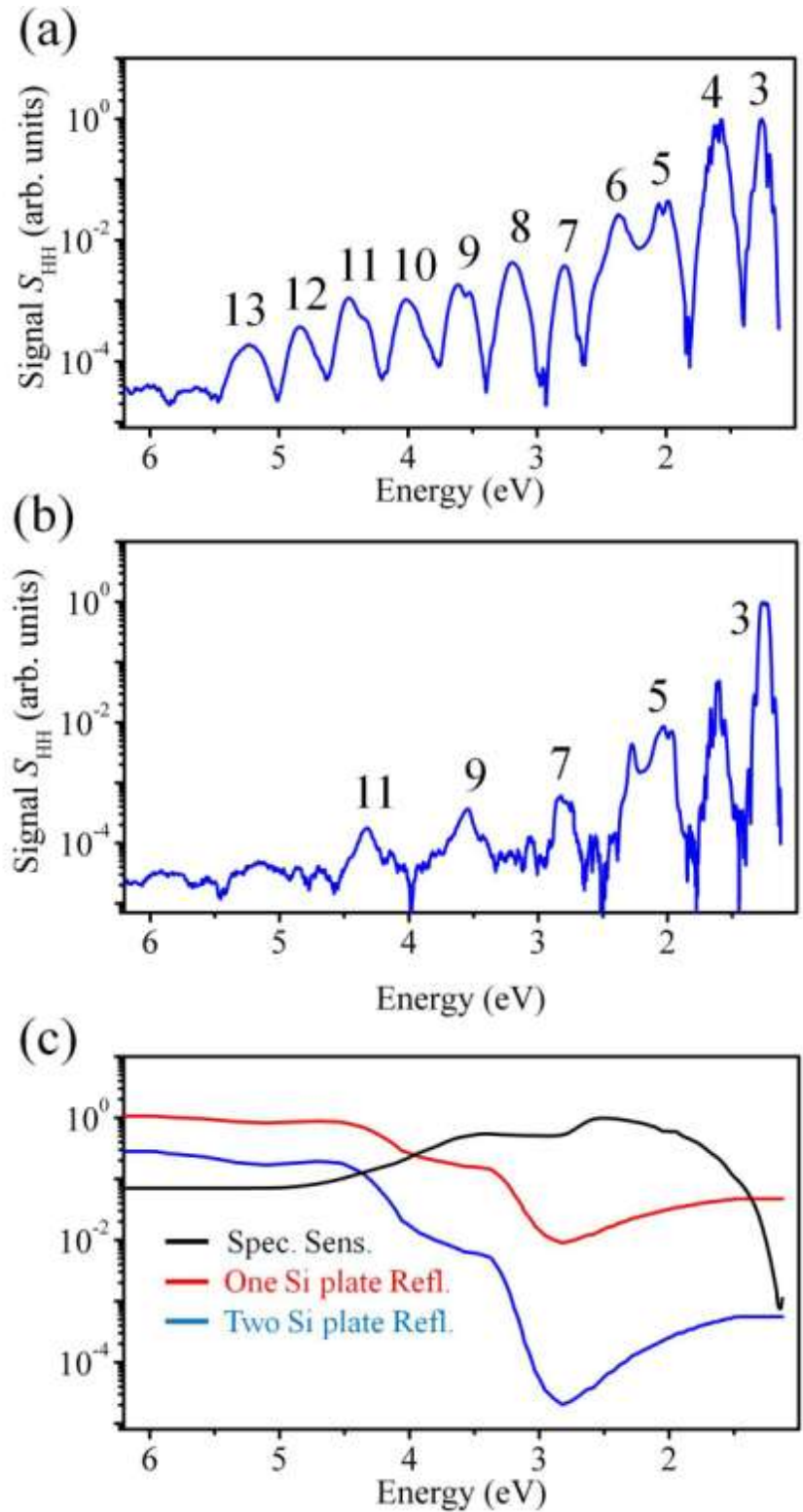

Fig. S4. (a) and (b) Harmonic spectra as were recorded by the conventional spectrometer when the optical axis of the crystal was parallel and perpendicular to the laser polarization, respectively. (c) The black, red and blue lines show the spectral response of the spectrometer, the reflectivity of one Si plate and the reflectivity of two Si plates, respectively.

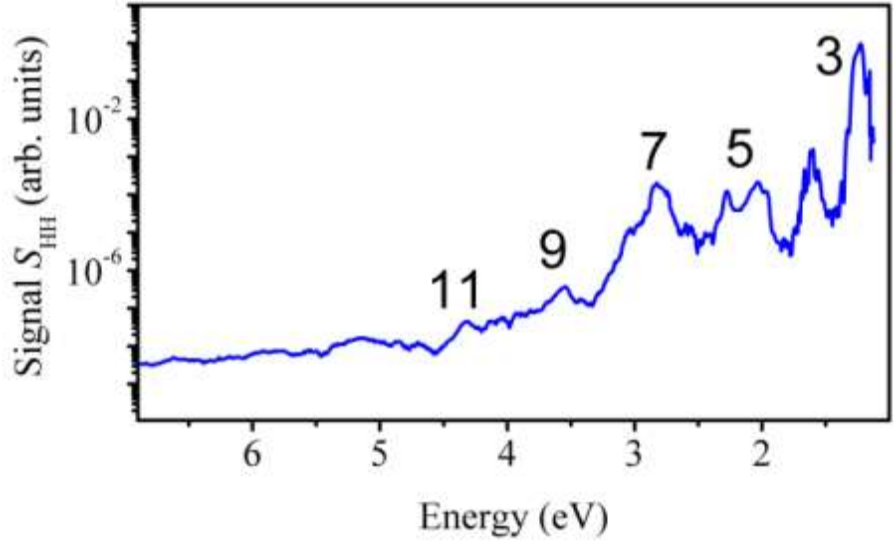

Fig. S5. High-Harmonic spectrum at the exit of the crystal recorded for $I_{IR_0} \approx 5 \times 10^{11}$ W cm$^{-2}$ when the optical axis of the crystal was perpendicular to the laser polarization.